\documentclass[aps,prl,showpacs,superscriptaddress,groupedaddress,twocolumn]{revtex4}
\usepackage{amsfonts}
\usepackage{amsmath}
\usepackage{mathrsfs}
\usepackage{amssymb}
\usepackage{graphicx}
\usepackage{dcolumn}
\usepackage{bm}
\usepackage{color}

\begin{document}
\title{$\mathcal{PT}$-Symmetric Phonon Laser}

\author{H. Jing}
\affiliation{CEMS, RIKEN, Saitama, 351-0198, Japan}
\affiliation{Department of Physics, Henan Normal University,
Xinxiang 453007, P.R. China}
\author{S. K. \"Ozdemir}
\affiliation{CEMS, RIKEN, Saitama, 351-0198, Japan}
\affiliation{Electrical and Systems Engineering, Washington
University, St. Louis, Missouri 63130, USA}
\author{Xin-You L\"{u}}
\affiliation{CEMS, RIKEN, Saitama, 351-0198, Japan}
\author{Jing Zhang}
\affiliation{CEMS, RIKEN, Saitama, 351-0198, Japan}
\affiliation{Department of Automation, Tsinghua University,
Beijing 100084, P.R. China}
\author{Lan Yang}
\affiliation{Electrical and Systems Engineering, Washington
University, St. Louis, Missouri 63130, USA}
\author{Franco Nori}
\affiliation{CEMS, RIKEN, Saitama, 351-0198, Japan}
\affiliation{Physics Department, University of Michigan, Ann
Arbor, MI 48109-1040, USA}
\date{\today}

\begin{abstract}
  By exploiting recent developments associated with coupled microcavities, we introduce the concept of $\mathcal{PT}$-symmetric phonon laser with balanced gain
  and loss. This is
 accomplished by introducing {gain} to one of the microcavities such that it balances the passive loss of the other. In the vicinity of the
  gain-loss balance, a strong nonlinear relation emerges between the intracavity photon intensity and the input power. This then leads to a giant enhancement
  of both optical pressure and mechanical gain, resulting in a highly efficient phonon-lasing action. These results provide a promising approach
  for manipulating optomechanical systems through $\mathcal{PT}$-symmetric concepts. Potential applications range from enhancing mechanical cooling to designing phonon-laser amplifiers.
\end{abstract}
 \maketitle

Recent advances in materials science and nanofabrication have led
to spectacular achievements in cooling classical mechanical
objects into the subtle quantum regime
(e.g.,\,\cite{OM-review1,cooling2,cooling3,cooling4}). These
results are having a profound impact on a wide range of research
topics, from probing basic rules of classical-to-quantum
transitions \cite{cooling4,strongOM,squeez,quantOM} {to creating}
novel devices operating in the quantum regime, e.g. ultra-weak
force sensors \cite{sensor} or electric-to-optical wave
transducers \cite{Dong,Lehnert}. The emerging field of cavity
optomechanics (COM) \cite{OM-review1} is also experiencing rapid
evolution { that is driven by studies aimed at understanding the
underlying physics and by the fabrication of novel structures and
devices enabled by recent developments in nanotechnology.}

The basic COM system includes a single resonator, where a
highly-efficient energy transfer between the mechanical mode and
intracavity photons is enabled by detuning an input laser from the
cavity resonance \cite{OM-review1}. A new extension, closely
related to the present study, is the {\it photonic molecule} or
compound microresonators \cite{compound,phononlaser1,Fan}, where a
tunable optical tunneling can be exploited to bypass the frequency
detuning requirement \cite{Fan}. More strikingly, in this
architecture, an analogue of two-level optical laser is provided
by phonon-mediated transitions between two optical supermodes
\cite{phononlaser1}. This phonon laser
\cite{phononlaser1,phononlaser2} provides the core technology to
integrate coherent phonon sources, detectors, and waveguides ---
allowing the study of nonlinear phononics \cite{qubit} and the
operation of functional phononic devices \cite{qubit2}.

In parallel to these works, intense interest has also emerged
recently in $\mathcal{PT}$-symmetric optics \cite{PT1,PT2,PT3}. A
variety of optical structures, whose behaviors can be described by
parity-time ($\mathcal{PT}$) symmetric Hamiltonians, have been
fabricated \cite{PT1}. These exotic structures provide
unconventional and previously-unattainable control of light
\cite{PT2,PT3,Sahin1,Sahin2}. In very recent work, by manipulating
the gain (in one active or externally-pumped resonator) to loss
(in the other, passive, one) ratio, Ref.\,\cite{Sahin1} realized
an optical compound structure with remarkable
$\mathcal{PT}$-symmetric features, e.g. field localization in the
active resonator and accompanied enhancement of optical
nonlinearity leading to nonreciprocal light transmission. However,
COM properties underlying the phonon-laser action in the
$\mathcal{PT}$-symmetric regime, where gain and loss are balanced,
remain largely unexplored.

Here we study a $\mathcal{PT}$-symmetric COM system which is
formed by two coupled microcavities, one of which has passive loss
(passive resonator $\mathbb{R}_\gamma$: no optical gain) and the
other has optical gain (active resonator $\mathbb{R}_\kappa$)
balancing the loss of $\mathbb{R}_\gamma$. In contrast to passive
COM, with single or coupled passive resonators,
$\mathcal{PT}$-symmetric COM features a transition from linear to
nonlinear regimes for intracavity photon intensity, by controlling
the gain-loss ratio. In this nonlinear regime, a giant enhancement
of both optical pressures and mechanical gain can be realized.
Consequently, in the $\mathcal{PT}$-symmetric regime, an
ultralow-threshold phonon laser is achievable by approaching the
gain-loss balance. We note that the phonon lasing action, through
energy exchange of two nondegenerate optical supermodes, exists
only in the $\mathcal{PT}$-symmetric regime. The enhanced
nonlinearity is responsible {for} the ultralow threshold of the
phonon laser and can also be useful for studying a wide variety of
optomechanical processes, e.g.,\, single-photon COM \cite{Rabl1}
or phononic mixing \cite{mixing,qubit2}. All relevant parameters
and techniques are well within the reach of current experimental
capabilities.

We consider two coupled whispering-gallery-mode (WGM) microtoroid
resonators. One of the microtoroids is fabricated from silica and
has passive loss (passive resonator $\mathbb{R}_\gamma$), whereas
the other microtoroid is fabricated from silica doped with
Er$^{3+}$ ions (active resonator $\mathbb{R}_\kappa$). Er$^{3+}$
ions emit photons in the $1550\,\mathrm{nm}$ band when
$\mathbb{R}_\kappa$ is optically pumped with a light in the
$1460\,\mathrm{nm}$ band. This provides the optical gain $\kappa$
to compensate for the optical losses and to amplify weak signal
light in the $1550\,\mathrm{nm}$ band \cite{Sahin1}. Evanescent
coupling between the two resonators exists only in the
$1550\,\mathrm{nm}$ band, assuring that the light in the
$1460\,\mathrm{nm}$ band only resides in $\mathbb{R}_\kappa$
\cite{Sahin1}. The amplified light in the $1550\,\mathrm{nm}$ band
then serves as the pump for a mechanical mode (frequency
$\omega_m$ and effective mass $m$) contained in the passive
resonator $\mathbb{R}_\gamma$ \cite{phononlaser1}. In order to
couple external light into and out of the WGMs, each microtoroid
is coupled to a different tapered-fiber waveguide (see Fig. 1).

By achieving population inversion in this system, the stimulated
emission of phonons can occur, in close analogy to an optical
laser. A phonon laser operating with threshold power $\sim7
\,\mu\mathrm{W}$ has been already demonstrated \cite{phononlaser1}
with two passive resonators. Extending this system to involve gain
and loss yields the Heisenberg equations of motion ($\hbar=1$)
\begin{align}
  &\dot a_1=\kappa a_1+iJa_2+\sqrt{2\kappa}a^{\mathrm{in}},\label{eq1} \\
  &\dot a_2=-\gamma a_2+iJa_1+iga_2x, \label{eq2}\\
  &\ddot{x}+\Gamma_m\dot
x+\omega^2_mx=\frac{g}{m}a^\dag_2a_2+\frac{\varepsilon^{\mathrm{in}}}{m},\label{eq3}
\end{align}
where $x=x_0(b+b^\dag)$ is the mechanical position operator with
$x_0=(2m\omega_m)^{-1/2}$ and the operator $b$ denoting the phonon
mode, {$g={\omega_c}/{R}$ is the COM coupling coefficient,} $J$ is
the inter-cavity coupling rate, $R$ is the microtoroid radius, and
$\gamma={\omega_c}/{Q_c}$ is the optical loss. The underlying
resonances are assumed degenerate with frequency
$\omega_c={c}/{\lambda}$, while $a_{1,2}$ and $b$ denote the
lowering operators for the optical and mechanical modes,
respectively. $\Gamma_m$ is the mechanical damping rate and the
operator $\varepsilon^{\mathrm{in}}$ describes the thermal
Brownian noise resulting from the coupling of the resonators to
the environment. The operator $a^{\mathrm{in}}$ is the optical
noise operator describing the signal light incident on
$\mathbb{R}_\kappa$; its mean value $f_{\mathrm{in}}=\langle
a^{\mathrm{in}}(t)\rangle$ is positive, and its fluctuations
denoted by $\delta a^{\mathrm{in}}(t)$ are assumed to be
delta-correlated \cite{PT3}, adding vacuum noise to the resonator
modes
\begin{align}
\langle \delta a^{\mathrm{in},\dag}(t)\delta
a^{\mathrm{in}}(t')\rangle=\delta(t-t'),~~\langle \delta
a^{\mathrm{in}}(t)\delta
a^{\mathrm{in},\dag}(t')\rangle=0.\nonumber
\end{align}

\begin{figure}[ht]
\includegraphics[width=8cm]{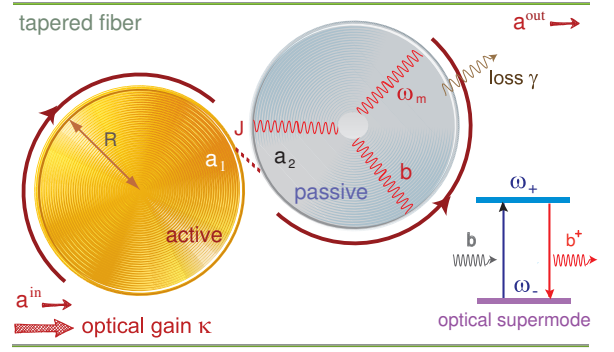}
\caption{(Color online) Gain-enhanced {optomechanics in} compound
whispering-gallery resonators. The optical tunnelling rate $J$ {is
tuned by changing the distance} between $\mathbb{R}_\kappa$ and
$\mathbb{R}_\gamma$. The corresponding {optical} supermodes
coupled by phonons are also plotted.} \label{fig2}
\end{figure}

By setting the time derivatives in Eqs.(\ref{eq1})-(\ref{eq3}) to
zero, we find the steady state of the dynamical variables as
\begin{align}
&a_{1,s}=\frac{-\sqrt{2\kappa}f_{\mathrm{in}}}{\kappa-J^2/(\gamma-igx_s)},\nonumber\\
&a_{2,s}=\frac{-iJ\sqrt{2\kappa}f_{\mathrm{in}}}{\kappa\gamma-J^2-i\kappa
gx_s},~~x_s=\frac{g|a_{2,s}|^2}{m\omega^2_m}.
\end{align}

The expressions of $a_{2,s}$ and $x_s$ can be combined to give
\begin{equation}
\frac{2\kappa J^2P_{\mathrm{in}}/R}{(\kappa\gamma-J^2)^2+(\kappa
gx_s)^2}=m\omega^2_m x_s,\label{eq5}
\end{equation}
which shows the balance of the radiation and spring forces. Here
we have used $P_{\mathrm{in}}=\omega_c|f_{\mathrm{in}}|^2$, with
$P_{\mathrm{in}}$ denoting the power of the signal light incident
on $\mathbb{R}_\kappa$. The cubic equation about $x_s$
characterizes the occurrence of bistability at higher input power
$P_{\mathrm{in}}$, as in passive COM systems \cite{X,bistability}.

If the coupling rate, gain and loss satisfy the condition $J^2={\kappa\gamma }$, there exists only one solution that is always positive
\begin{equation}
a_{2,s}=\frac{\sqrt{2\gamma}f_{\mathrm{in}}}{gx_s},~~x_s=\left({\frac{2P_{\mathrm{in}}R}{m\omega^2_m\omega_cQ_c}}\right)^{1/3},
\end{equation}
for $any$ value of the input power $P_{\mathrm{in}}$. In contrast,
even below the onset of bistability ($gx_{s}\ll \gamma$), passive
COM systems have very different results
$$|a_{1,s}|_\mathrm{p}=|a_{2,s}|_\mathrm{p}=\frac{f_{\mathrm{in}}}{\sqrt{2\gamma}},~~x_{s,\mathrm{p}}=\frac{P_{\mathrm{in}}}{2\gamma m\omega^2_{m}R},$$
where for comparison we take $J\sim\gamma$, and the subscript
$\mathrm{p}$ denotes the COM system with coupled passive
resonators. At the {exact} gain-loss balance i.e.
$\delta\equiv\kappa/\gamma=1$, the ratio of steady-state
populations in the passive resonator $\mathbb{R}_\gamma$ for
$\mathcal{PT}$-symmetric and passive COM systems is {given by}
\begin{equation}
\eta \equiv \frac{|a_{2,s}|^2}{|a_{2,s}|^2_\mathrm{p}}=
\frac{x_s}{x_{s,\mathrm{p}}}=\Big(\frac{4\gamma^2m\omega^2_mR^2}
{\omega_cP_{\mathrm{in}}}\Big)^{2/3}.
\end{equation}
As in relevant experiments \cite{Sahin1,phononlaser1}, the
parameter values are taken as $\lambda=1550\,\mathrm{nm}$,
$Q_c=3\times 10^7$, $R\sim 34.5\,\mu\mathrm{m}$, and
$\omega_m=2\pi\times 23.4\,\mathrm{MHZ}$, $m=5\times
10^{-11}\,\mathrm{Kg}$, $2Q_m\sim Q_c/10^5$, which leads to
$g\sim5.61\,\mathrm{GHz/nm}$, $\gamma \sim 6.45\,\mathrm{MHz}$,
$\Gamma_m=\omega_m/Q_m \sim 2.4\times 10^5\,\mathrm{Hz}$. For
these values, $\Gamma_m/\gamma\ll 1$, {implying that the system is
well} within the phonon stimulated regime \cite{phononlaser1}. The
condition $gx_s\ll \gamma$ is fulfilled for $P_{\mathrm{in}}\ll
137\,\mu\mathrm{W}$.

A main feature of the {present work} is that, by approaching the
gain-loss balance both optical pressure and mechanical gain can be
significantly amplified. For $\delta=1$, an enhancement of two
orders of magnitude in the intracavity field intensity can be
achieved. Namely, $\eta \sim 106$ or $\sim 29.5$ for
$P_{\mathrm{in}}=1\,\mu\mathrm{W}$ or $7\,\mu\mathrm{W}$. The
$\mathcal{PT}$-symmetric COM system performs better than the
passive COM system with threshold power
$P_{\mathrm{th},\mathrm{p}}=7\,\mu\mathrm{W}$, even at
significantly lower input powers ($P_{\mathrm{in}}\ll
7\,\mu\mathrm{W}$) \cite{ZJ}. When the gain-loss ratio deviates
from the {exact balance} condition (e.g.,\,$\delta
>3$) or when $P_{\mathrm{in}}$ exceeds a threshold
$1.1\,\mathrm{mW}$, our system, built from coupled passive and
active resonators, behaves in the same way as a system of coupled
passive resonators.

Figure 2(a) shows the steady-state populations of intracavity
photons in the passive resonator $\mathbb{R}_\gamma$. For weak
input power ($P_{\mathrm{in}}\leq10\,\mu\mathrm{W}$), passive COM
systems feature linear responses \cite{OM-review1,bistability}, in
sharp contrast to the situation of balanced gain and loss, for
which significant optical nonlinearity and the accompanied giant
enhancement of COM interactions appear. This is strongly
reminiscent of the situation encountered in nonreciprocal wave
transmission in $\mathcal{PT}$-symmetric optical \cite{Sahin1} or
electric \cite{transport} structures. In this nonlinear regime,
the transmission rate is vastly different when driving the coupled
micro-resonators from the left or the right side
\cite{Sahin1,Rabl2}. {Similar features as in Fig.\,2 can be
observed by changing the optical coupling rate $J/\gamma$ at a
fixed gain-loss ratio $\kappa/\gamma$ (see the Supplemental
Material \cite{ZJ}).}

\begin{figure}[ht]
\includegraphics[width=8cm]{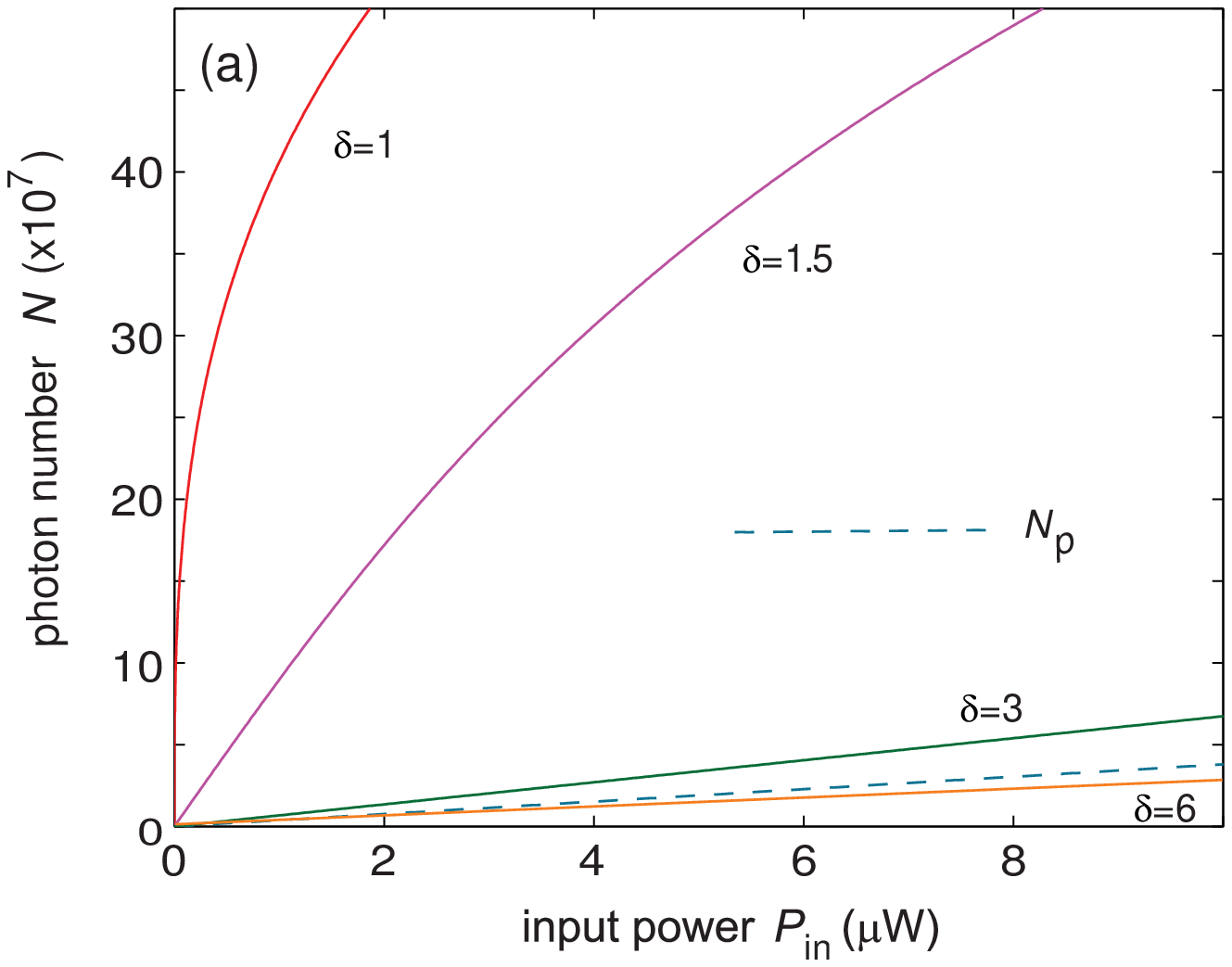}
\includegraphics[width=8cm]{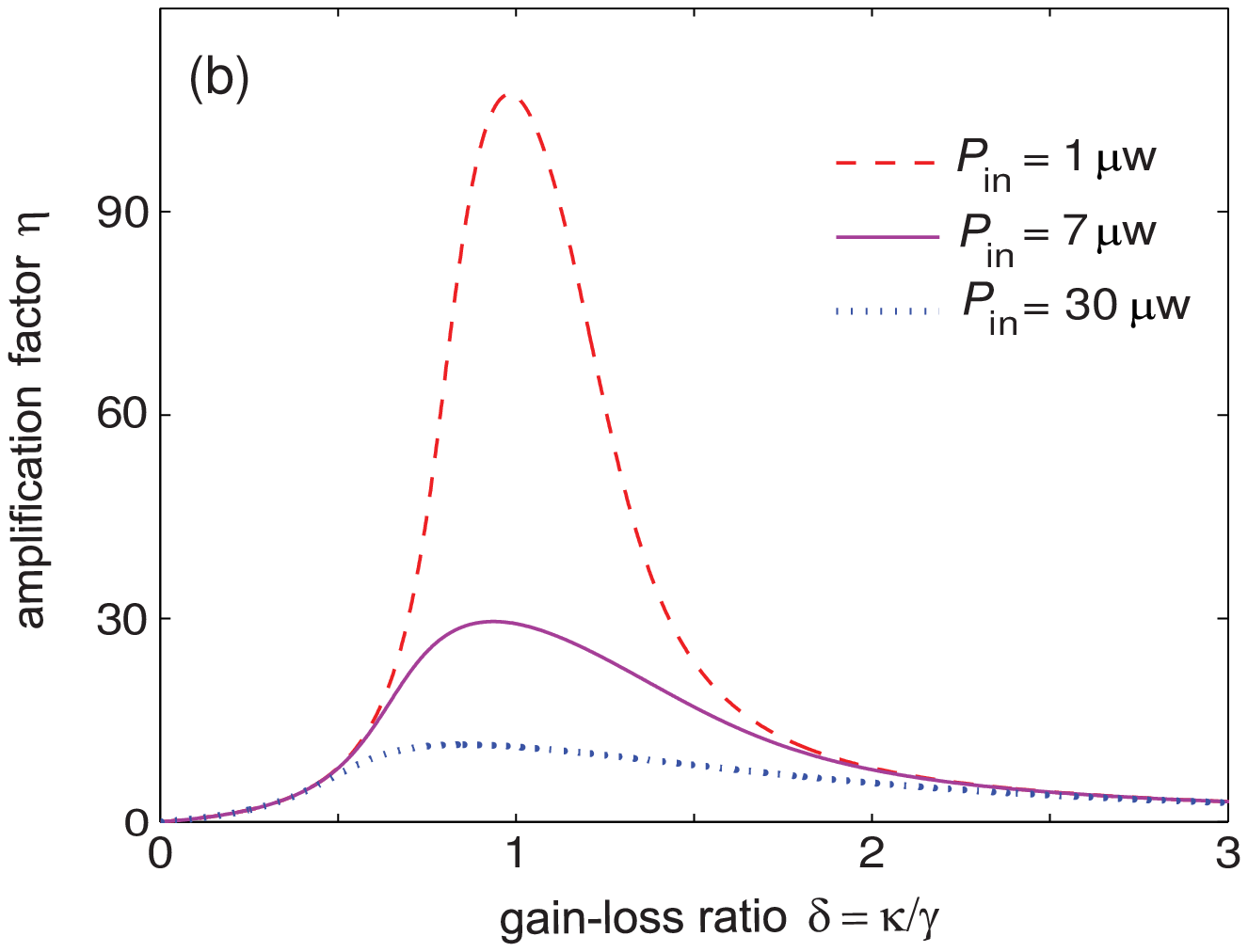}
\caption{(Color online) (a) Steady-state cavity-photon number $N$:
for $P_{\mathrm{in}}<30\,\mu\mathrm{W}$, the passive system shows
linear responses (dashed line); in contrast, nonlinear behavior
emerges for the gain-loss balanced system ($\delta=1$); (b)
optomechanical amplification factor $\eta$ with tunable ratio
$\delta\equiv\kappa/\gamma$.} \label{fig2}
\end{figure}

Figure 2(b) shows the associated optical amplification factor
$\eta(\delta, P_{\mathrm{in}})$, featuring resonance peaks for
gain-loss balance. This indicates that for ultraweak input light
($P_{\mathrm{in}}\rightarrow 0$), in contrast to the present
structure, the intracavity photon number approaches zero very
rapidly for passive COM systems. {At a fixed value of
$P_{\mathrm{in}}$, the radiation pressure in $\mathbb{R}_\gamma$
containing the mechanical mode can be significantly enhanced by
tuning only the gain-loss ratio in the $\mathcal{PT}$-symmetric
system.} This is also reminiscent of resonantly-enhanced light
transmissions in $\mathcal{PT}$-symmetric optical structures
\cite{Sahin1}.

The gain-enhanced nonlinearity is potentially useful in a wide
range of phononic engineering systems, e.g.\,stiffer trapping and
further cooling a mechanical object deep into its quantum ground
state \cite{OM-review1,cooling2,cooling3,cooling4}. Instead of
this, here we study its impact on coherent phonon lasing.
According to Grudinin $et~al.$ \cite{phononlaser1}, compound
resonators provide a phonon analog of a two-level laser by
replacing the photon-mediated electronic transitions with
phonon-mediated optical transitions. The optical inversion then
produces coherent mechanical gain at a breathing mode with
frequency $\omega_m$, leading to a phonon laser above the
threshold power $P_{\mathrm{th},\mathrm{p}}\sim 7\,\mu\mathrm{W}$
\cite{phononlaser1}.

The Hamiltonian of the multi-mode COM system was already given
elsewhere \cite{compound}. In the rotating frame at the signal
laser frequency $\omega_L$, the interaction term can be mapped
into a simple form, i.e.
\begin{equation}
H_{\mathrm{int}}=-J(a_1^\dag a_2+h.c.)-ga_2^\dag a_2 x
~\longrightarrow \frac{gx_0}{2}(pb^\dag+bp^\dag),
\end{equation}
where the optical inversion operator $p=a_-^\dag a_+$ is defined
with the supermode operator $a_\pm=({a_1\pm a_2})/{\sqrt{2}}$.
Then we have the equations of motion
\begin{align}
\dot{b}=&\left(-\Gamma_m-i\omega_m \right)b-i\frac{gx_0}{2}p, \\
\dot{p}=&\left(\kappa-\gamma-i\Delta\omega\right)p-i\frac{gx_0\Delta
n}{2}b,
\end{align}
where $\Delta n=n_+-n_-,$ $\Delta \omega=\omega_+-\omega_-,$ with
$n_\pm$ or $\omega_\pm$ being the density or frequency of the
supermodes. Here, unlike passive COM systems
\cite{phononlaser1,Fan,compound}, the presence of active gain
changes both the mode splitting and linewidth of the supermodes,
i.e.
\begin{align}
\omega_{\pm}=&\omega_c
\pm
\left[{J^2-\left({\kappa+\gamma}\right)^2}/4\right]^{1/2} ~~\Rightarrow ~~\Delta\omega\neq 0,\nonumber\\
\gamma_\pm=&(\kappa-\gamma)/2,
\end{align}
for strong optical tunnelling rate, i.e.\,$J\geq
(\kappa+\gamma)/2$, a so-called unbroken-$\mathcal{PT}$-symmetry
regime has been identified experimentally in a purely optical
structure \cite{Sahin1}. Only in this regime the supermodes can be
distributed evenly across the resonators and hence enabling the
compensation of loss with gain. In contrast, for the
broken-$\mathcal{PT}$-symmetry regime with weaker inter-cavity
coupling $J<(\kappa+\gamma)/2$, no supermode splitting exists at
all \cite{ZJ},
\begin{align}
\omega_{\pm}=&\omega_c ~~~\Rightarrow ~~~\Delta\omega=0,\nonumber\\
\gamma_{\pm}=&(\kappa-\gamma)/{2}\pm\left[\left(\kappa+\gamma\right)^2/4-J^2\right]^{1/2}.
\end{align}

This leads to {the following important result}: {the phonon lasing
exists only in the $\mathcal{PT}$-symmetric regime}, where the
optical supermodes are non-degenerate and thus can exchange energy
through the phonon mode. In contrast, no exchange channel exists
at all in the $\mathcal{PT}$-broken regime where supermodes become
spontaneously localized in either the amplifying or {the} lossy
resonator \cite{Sahin1}. {The presence of optical gain enables one
to drive the system between these two regimes on-demand,  as well
as to controllably set the spectral distance between the
supermodes using the interplay between the gain-loss ratio and the
intracavity-coupling strength.} This is very different from
passive COM systems, where the supermode splitting always exists
for $J>0$, i.e.\,$\Delta\omega=2J$ \cite{phononlaser1}. As a
signature of the first-order coherence, the stimulated emission
linewidth is much narrower in comparison with that below the
lasing threshold \cite{phononlaser1}.

Solving Eq.\,(10) in the frequency domain gives
\begin{equation}
p[\omega]=i(gx_0\Delta n/2)\mathcal{C}[\omega]b,
\end{equation}
with a cavity factor
\begin{equation}
\mathcal{C}[\omega]=\left[\kappa-\gamma+i(\omega-\Delta\omega)\right]^{-1}.
\end{equation}
Inserting this solution into Eq.\,(9) gives
\begin{equation}
\dot{b}=\left[G-\Gamma_m-i\omega_m-i(gx_0/2)^2(\omega-\Delta\omega)|\mathcal{C}|^2\Delta
n\right]b,
\end{equation}
where the mechanical {gain $G$ is given by}
\begin{eqnarray}
G=\frac{\left(gx_0/2\right)^2n_+\left(\kappa-\gamma\right)}{\left(\omega_+-\omega_--\omega_m\right)^2+\left(\kappa-\gamma\right)^2},
\end{eqnarray}
for the blue-supermode threshold density of $n_+$ at the line
center.

\begin{figure}[ht]
\includegraphics[width=8.2cm]{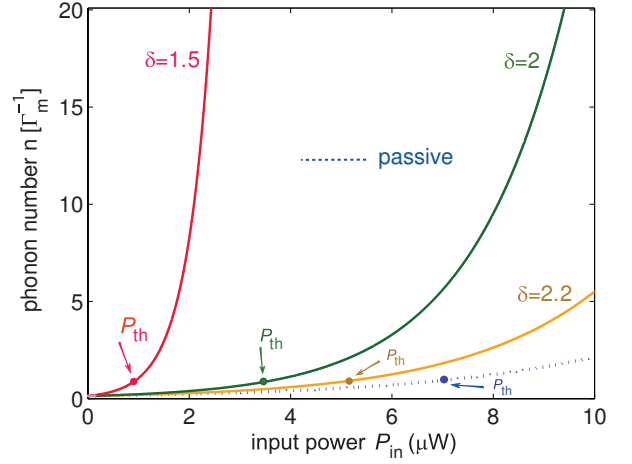}
\caption{(Color online) Plot of the stimulated emitted phonon
number $n[\Gamma_m^{-1}]=\exp[2(G-\Gamma_m)/\Gamma_m]=\exp[2(\beta
P_{\mathrm{in}}/\omega_+-1)]$ in the $\mathcal{PT}$-symmetric or
the passive system, as a function of $P_{\mathrm{in}}$. Here
$\beta\equiv(gx_0|\mathcal{C}|/2)^2$, and to compare with
Ref.\,\cite{phononlaser1}, we choose $2J=\omega_m$ (well within
the $\mathcal{PT}$-symmetric regime). The threshold power
$P_{\mathrm{th}}$ denoted by the thick points is obtained for
$G=\Gamma_m$. The threshold value of the passive COM system is
$\sim 7\,\mu \mathrm{W}$, agreeing well with the experiment
\cite{phononlaser1}, which can be significantly lowered for
$\delta\rightarrow 1$.  } \label{fig2}
\end{figure}

The number of emitted phonons is determined by the threshold
condition $G=\Gamma_m$ (see Fig. 3). This condition, together with
$P_{\mathrm{th}}=n_+(\gamma_-+\gamma_+)\omega_+$, yields
\begin{equation}
P_{\mathrm{th}}=4\,\Gamma_m\omega_+{\left[\left(\omega_+-\omega_--\omega_m\right)^2+\left(\kappa-\gamma\right)^2\right]}/(gx_0)^2.
\end{equation}
Clearly $G\rightarrow\infty$, $P_{\mathrm{th}}\rightarrow 0$ under
the following conditions:
\begin{align}
&\mathrm{(i)}~~\omega_m=2\left[J^2-\left(\kappa+\gamma\right)^2/4\right]^{1/2},\nonumber\\
&\mathrm{(ii)}~~ \kappa=\gamma
~~\mathrm{(gain}\mathcal{-}\mathrm{loss~balance)},
\end{align}
indicating an ultralow-threshold phonon laser by tuning both
$\kappa/\gamma$ and $J/\gamma$ (in the $\mathcal{PT}$-symmetric
regime), which is otherwise unattainable for passive COM devices.
The tunability of these parameters was already demonstrated in a
very recent experiment \cite{Sahin1}. With the parameter values
given above, the threshold power of the passive COM system can be
estimated as $P_{cT,\mathrm{p}}\sim\,7\,\mu\mathrm{W}$, agreeing
well with the experiment \cite{phononlaser1}. In contrast, by
compensating loss with gain ($\delta\rightarrow 1$), the
phonon-lasing threshold can be significantly lowered, also in
agreement with the resonant enhancement shown in Fig. 2. We see
that the two tunable parameters (the optical tunnelling rate and
the gain-loss ratio), when acting in concert, provide more
flexibility in COM control.

Finally, note that the presence of optical gain provides an
additional degree of freedom to control the dynamics of the COM
system. Namely, tuning the gain-loss ratio can switch between
stable and bistable operations (see the Supplemental Materials
\cite{ZJ}). A similar situation can also occur in electronic
circuits \cite{transport}.

In summary, {we have studied} a compound-resonator COM system in
the presence of active gain {in one resonator and passive loss in
the other.} At the gain-loss balance, an optical nonlinearity is
observable for steady-state populations {even for ultraweak input
powers.} This gain-induced nonlinearity leads to a giant
enhancement of both optical pressure and mechanical gain, enabling
then an ultralow-threshold phonon laser in the
$\mathcal{PT}$-symmetric regime. As demonstrated by a recent
experiment \cite{phononlaser1}, increasing the optical tunnelling
rate can lead to a transition from broken to unbroken
$\mathcal{PT}$-symmetric regimes, i.e.\,realizing not only the
exchange of two subsystems but also changing gain to loss and vice
versa. The linear-to-nonlinear transition, i.e.\,the giant
enhancement of the intracavity field intensity and then the
mechanical gain, can be realized when approaching the gain-loss
balance. We stress {that $\kappa/\gamma\rightarrow 1$ and $J\geq
(\kappa+\gamma)/2$ should be satisfied simultaneously} to observe
the unidirectional propagation of light in a purely optical
experiment \cite{Sahin1}, and now {to obtain} an
ultralow-threshold phonon laser in a COM system.

Our work opens up exciting new perspectives for COM control with
unconventional $\mathcal{PT}$-structures \cite{PT2,PT3,PT1}, e.g.,
asymmetric wave transport \cite{transport}, gain-enhanced
mechanical mixing \cite{mixing}, ultraslow light in
$\mathcal{PT}$-structures \cite{Fan,slowlight}, and
$\mathcal{PT}$-enhanced phononic squeezing or entanglement
\cite{squeez}. In future works we also plan to study a
$\mathrm{CW}$ phonon laser with a three-level system
\cite{phononlaser2}, a multi-mode phonon laser with a tunable
gain-loss ratio, or a triple-resonator set-up
\cite{3mirror,mixing} with two active elements on both sides of a
mechanical mode. In view of rapid advances in compound
micro-structures \cite{compound,Fan,Sahin1,Sahin2,phononlaser1},
particularly those on phonon lasers \cite{phononlaser1} and
$\mathcal{PT}$-symmetric resonators \cite{Sahin1}, we believe that
these studies will be highly accessible in experiments in the near
future.

HJ thanks Y. Nakamura, K. Usami, and R. Yamazaki for helpful
discussions. HJ is supported by the NSFC under Grant No.11274098
and the Henan Excellence Plan (T555-1202). SKO and LY are
supported by ARO Grant No. W911NF-12-1-0026 and the NSFC under
Grant No. 61328502. XYL is supported by NSFC under Grant No.
11374116. JZ is supported by the NSFC under Grant Nos. 61174084,
61134008, and the NBRPC (973 Program) under Grant No.
2014CB921401. FN is supported by the RIKEN iTHES Project, MURI
Center for Dynamic Magneto-Optics, and a Grant-in-Aid for
Scientific Research\,(S).

\section*{Supplementary Materials for ``$\mathcal{PT}$-Symmetric
Phonon Laser''}

This document consists of three parts: (I) amplification factor
with varying the optical tunnelling rate $J$;  (II) supermode
splitting and linewidth; and (III) stability analysis with the
tunable gain-loss ratio $\delta$.

\subsection*{I.~Amplification factor with varying $J$}

Figure 2 in the main text shows the steady-state populations of
intracavity photons in the passive resonator, by numerically
solving Eq.\,(5). A notable feature of the
$\mathcal{PT}$-symmetric COM system is the emerging resonance
peaks of the optical amplification factor $\eta$ around the
gain-loss balance $\delta=1$ (see Fig.\,2b), where we choose to
change the values of gain-loss ratio $\delta$ but with fixed
optical tunnelling rate ($J/\gamma=1$). Here we show in Fig.\,S1
that for fixed $\delta$, similar features can be observed by
changing the optical tunnelling rate $J/\gamma$.

\begin{figure}[ht]
\includegraphics[width=8cm]{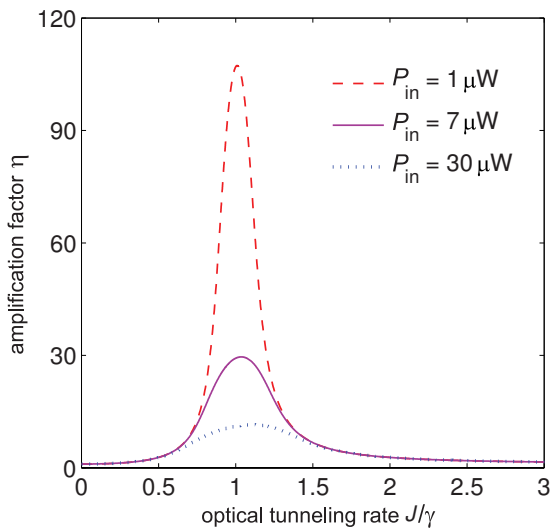}
\caption{(Color online) Optomechanical amplification factor $\eta$
versus the tunable optical tunnelling rate $J/\gamma$, for the
fixed value of gain-loss ratio $\delta=1$. Here the amplification
factor $\eta$ is by the Eq.\,(7) of the main text (having the
$\mathcal{PT}$-symmetric result divided by the result for the
passive COM case).} \label{fig2}
\end{figure}

\begin{figure}[ht]
\includegraphics[width=8.5cm]{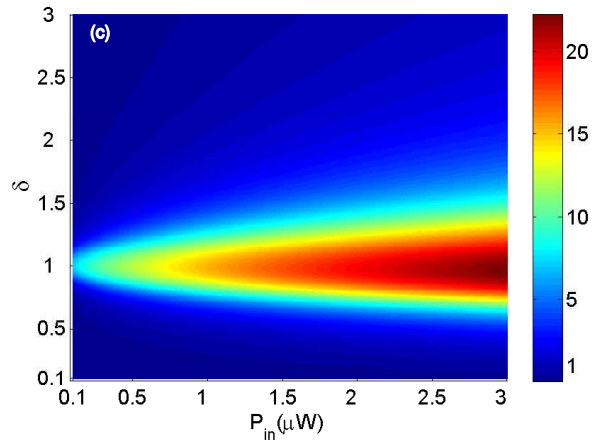}
\caption{(Color online) Relative amplification factor $\xi$ versus
tunable gain-loss ratio $\delta\equiv\kappa/\gamma$ and the input
power $P_\mathrm{in}$. The vertical color bar refers to the values
of $\xi$.}
\end{figure}

Even when comparing with a passive COM system with threshold power
$P_{\mathrm{th},\mathrm{p}}=7\,\mu\mathrm{W}$, the
$\mathcal{PT}$-symmetric COM system performs better for
significantly lower input power, i.e.
$$\xi\equiv \frac{x_s(\delta,
P_{\mathrm{in}})}{x_{s,\mathrm{p}}(P_{\mathrm{in}}=P_{\mathrm{th},\mathrm{p}})}\geq
1,$$ which, for $\delta=1$, can be realized for
$P_{\mathrm{in}}\geq 3\times 10^{-4}\, \mu\mathrm{W}$. For
instance, $\xi\sim 15.9$ or $29.5$ for $P_{\mathrm{in}}=1
\,\mu\mathrm{W}$ or $7\,\mu\mathrm{W}$. Figure\,S2 plots
$\xi(\delta, P_{in})$, by comparing the $\mathcal{PT}$-symmetric
system working below the threshold $7 \,\mu\mathrm{W}$ and the
passive COM system working with $7\, \mu\mathrm{W}$. For
$\delta\rightarrow 1$, $P_{\mathrm{in}}>0.1 \,\mu\mathrm{W}$, the
enhancement effect is significant even in this situation.

\subsection*{II.~Supermode splitting and linewidth}

\begin{figure}[ht]
\includegraphics[width=8cm]{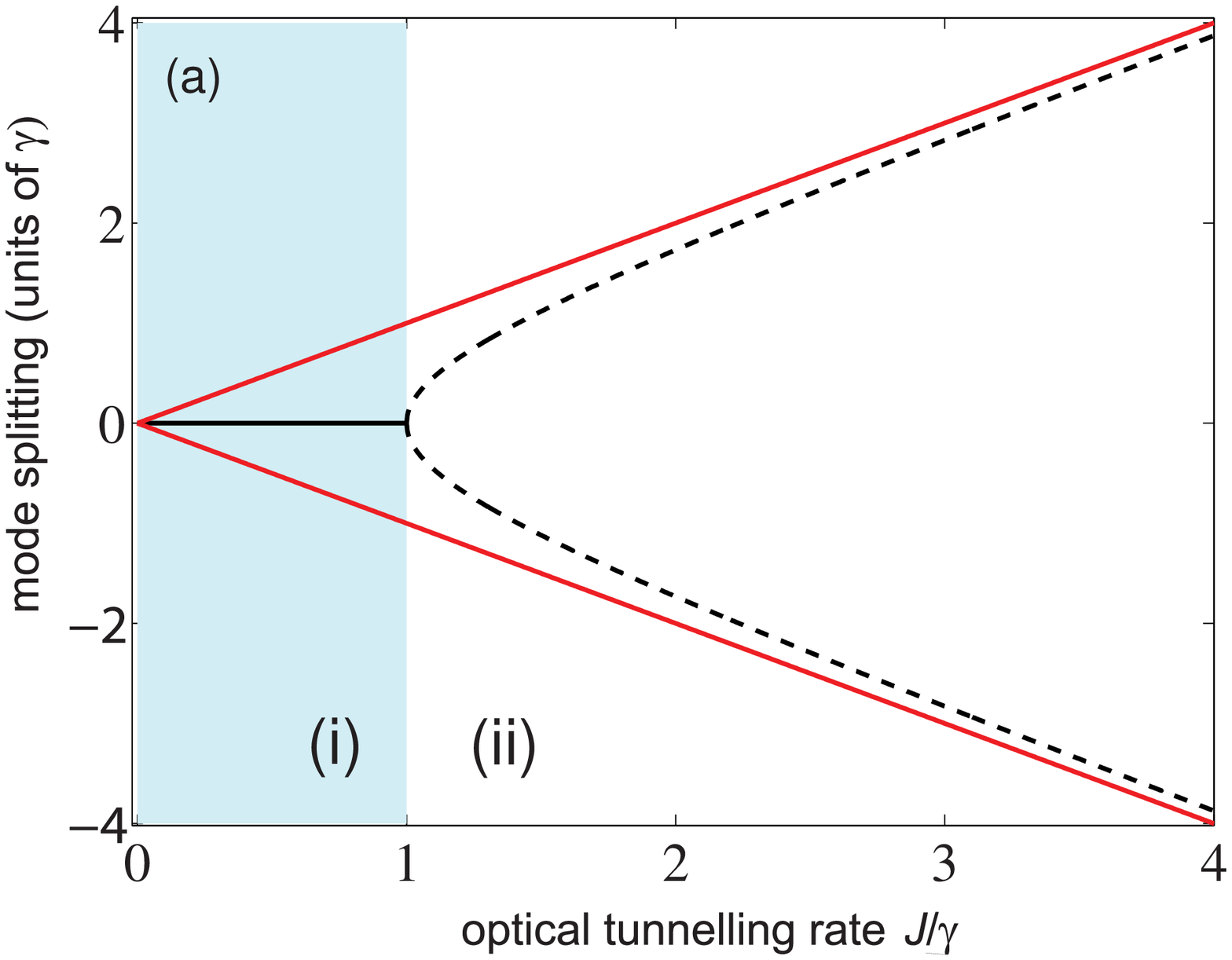}
\includegraphics[width=8cm]{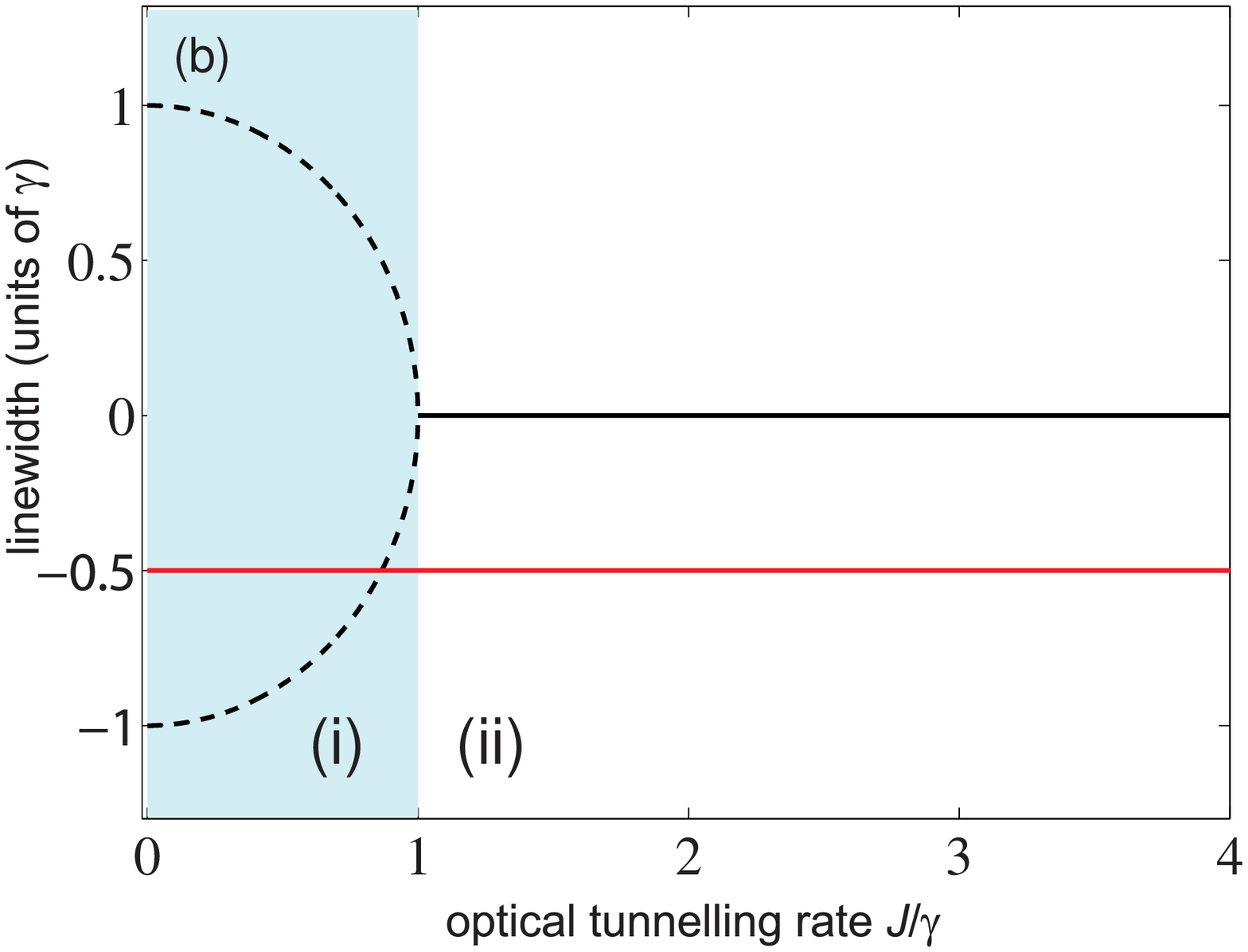}
\caption{(Color online) Mode-splitting (a) and linewidth (b) of
the supermodes in the $\mathcal{PT}$-symmetric COM system (black
curves) or the passive COM system (red lines), as a function of
$J/\gamma$ (see also Ref.\,\cite{Bender}). The $PT$-symmetry holds
in (ii) and not in (i).} \label{fig2}
\end{figure}

The non-Hermitian Hamiltonian of the system (comprising the
optical gain, the optical supermodes, and the phonon mode) can be
written at the simplest level as
\begin{eqnarray}
H_{\rm
tot}&=&\left(\omega_+-i\gamma_+\right)a_+^{\dagger}a_++\left(\omega_--i\gamma_-\right)a_-^{\dagger}a_-\nonumber\\
&&+\omega_m
b^{\dagger}b+\frac{gx_0}{2}\left(ba_+^{\dagger}a_-+a_-^{\dagger}a_+b^{\dagger}\right).
\end{eqnarray}
The weak driving terms are not explicitly shown here. The specific
expressions of $\omega_\pm$ and $\gamma_\pm$ are different in two
distinct regimes (see the main text): (i) the regime which was
identified as the broken-$\mathcal{PT}$-symmetry phase for a
purely optical structure \cite{Sahin1}, characterized by
$\left(\kappa+\gamma\right)/2>J$; (ii) the regime with strong
inter-cavity coupling $\left(\kappa+\gamma\right)/2\leq J$, which
for a purely optical system, was identified experimentally as the
unbroken-$\mathcal{PT}$-symmetry phase \cite{Sahin1}. We note that
only the supermodes with unbroken $\mathcal{PT}$-symmetry can be
distributed evenly across the coupled resonators, hence enabling
the compensation of loss with gain. In contrast to this, the
$\mathcal{PT}$-broken supermodes become spontaneously localized in
either the amplifying or lossy resonator, hence experiencing
either net gain or loss (see Ref.\,\cite{Bender} for more
details).

As Fig.\,S3 shows, the phonon lasing action can exist only in the
$\mathcal{PT}$-symmetric regime where the optical supermodes are
non-degenerate and thus can exchange energy through the phonon
mode. Figure S3 is similar to that as was shown in all
$\mathcal{PT}$-symmetric systems, e.g. in Ref. \cite{Sahin1};
nevertheless, here we show for the first time that the
ultralow-threshold phonon lasing can exist only in the
unbroken-symmetry regime, not in the broken-symmetry regime.

We stress that, by increasing the optical coupling rate $J$, one
can realize the transition from the broken to the unbroken
$\mathcal{PT}$-symmetric regimes. That is, realizing not only the
exchange of two subsystems (micro-resonators), but also changing
the gain to loss and vice versa. By tuning the gain-loss ratio,
one can realize the transition from linear to nonlinear regimes,
i.e.\,the giant enhancement of the intracavity field intensity and
then the mechanical gain. Both of these two conditions (strong $J$
and $\delta=1$) are required to observe the unidirectional wave
propagation in a purely optical system \cite{Sahin1} and now the
ultralow-threshold phonon laser in a COM system. It is the
presence of active gain which makes it possible to realize these
two conditions simultaneously.

\subsection*{III.~Stability analysis with tunable $\delta$}

Finally, we mention that in the vicinity of the gain-loss balance,
the stability properties of the COM system can also be
significantly modified. To see this we need to study the role of
thermal noise on the mechanical response. This is accomplished by
linearizing Eqs.\,(1-3) and then studying the fluctuations of the
operators. With the equations of motion as Eqs.\,(1-3) in the main
text, including also the optical detunings
$\Delta_i=\omega_{c,i}-\omega_L$ $(i=1,2)$ between the two
resonators and the input signal laser, the mean values of the
optical and mechanical modes then satisfy the following equation
\begin{eqnarray}
\left( \kappa -i\Delta _{1}\right) a_{1,s}-iJa_{2,s}-\sqrt{2\kappa }%
\Omega _{d} &=&0,   \\
\left( -\gamma -i\Delta _{2}\right) a_{2,s}-iga_{2,s}x-iJa_{1,s} &=&0,   \\
-m\omega _{m}^{2}x_{s}-g\left\vert a_{2,s}\right\vert ^{2} &=&0,
\end{eqnarray}%
with $\Omega _{d}=\sqrt{P_{\mathrm{in}}/\hbar \omega _c}$. From
these equations we can obtain the following polynomial for the
input power $P_{\mathrm{in}}$,
\begin{widetext}
\begin{equation}
P_{\mathrm{in}}=\frac{\omega _{c,1}}{2\kappa J^{2}}\left[
\begin{array}{c}
\frac{g^{4}}{m^{2}\omega _{m}^{4}}\left( \Delta
_{1}^{2}+\kappa ^{2}\right) N^{3}+\frac{g^{2}}{m\omega _{m}^{2}}%
\left( 2\gamma \Delta _{1}\kappa +2J^{2}\Delta _{1}-\Delta
_{1}^{2}\Delta
_{2}-2\gamma \kappa \Delta _{1}-2\kappa ^{2}\Delta _{2}\right) N^{2} \\
+\left( \gamma ^{2}\kappa ^{2}+J^{4}-2\gamma \kappa J^{2}+\Delta
_{1}^{2}\Delta _{2}^{2}+2\gamma \kappa \Delta _{1}\Delta
_{2}-2J^{2}\Delta _{1}\Delta _{2}+\gamma ^{2}\Delta
_{1}^{2}+\kappa ^{2}\Delta
_{2}^{2}-2\gamma \Delta _{1}\kappa \Delta _{2}\right) N%
\end{array}%
\right],
\end{equation}%
\end{widetext}
where $N$ denotes the photon number inside the passive resonator.
For a passive COM system, increasing the input power can lead to
unstable evolutions of the system. For our present active system,
however, a bistability-free regime is accessible by tuning
$\delta$, which is reminiscent of that in a
$\mathcal{PT}$-symmetric electronics system \cite{transport}.

\begin{figure}[ht]
\includegraphics[width=8.5cm]{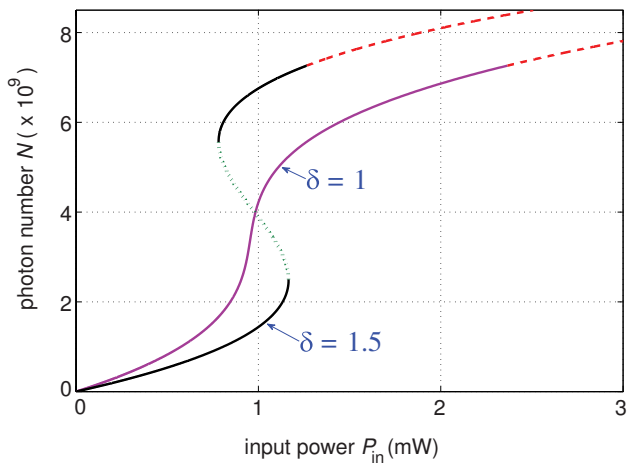}
\caption{(Color online) Mean population of photons in the passive
resonator containing the mechanical mode. The bistability feature
seen for $\delta=1.5$, which is confirmed to be similar to the
passive COM situation, can be removed at the gain-loss balance.
Dotted and dashed curves denote different types of instability
(see Ref.\,\cite{bistability}). All relevant parameter values are
given in the main text, including here also the optical detunings
$\Delta_1/\omega_m=0.03$ and $\Delta_2/\omega_m=0.15$.}
\label{fig2}
\end{figure}

Now by applying the expansions: $a_{1}=a_{1,s}+\delta a_{1},~~
a_{2}=a_{2,s}+\delta a_{2},~~x=x_s +\delta x,$ it is
straightforward to linearize Eqs.\,(1-3) to obtain
\begin{align}
&\frac{d\delta a_1}{dt}=\kappa\delta a_1-iJ\delta
a_2+\sqrt{2\kappa}\delta a^{\mathrm{in}},\nonumber \\
&\frac{d\delta a_2}{dt}=-\gamma\delta a_2-iJ\delta
a_1-iga_{2,s}\delta x-igx_s\delta
a_2, \\
&\frac{d^2\delta x}{dt^2}+\Gamma_m\frac{d\delta
x}{dt}+\omega_m^2\delta x=\frac{g}{m}\big(a_{2,s}^*\delta
a_2+a_{2,s}\delta
a^*_2\big)+\frac{\delta\varepsilon^{\mathrm{in}}}{m},\nonumber
\end{align}
for these fluctuations. The resulting solutions can be compactly
written as
$$
\delta x[\omega]=\chi[\omega]\,\delta\varepsilon^{\mathrm{in}},
$$
where the susceptibility is in the form
\begin{align}
\chi^{-1}[\omega]=m\left(\omega^2_m-\omega^2-i\omega\Gamma_m
+\frac{2g^2}{m}|a_{2,s}|^2\mathbf{Re}\mathbb{Y}[\omega]\right),
\end{align}
with a cavity factor
$$
\mathbb{Y}^{-1}[\omega]=-\omega-i\gamma+gx_s+\frac{J^2}{\omega-i\kappa},
$$
for a thermally-driven system. Hence, by tuning the gain-loss
ratio, COM properties (such as the mechanical susceptibility
\cite{X} and the bistability features) can be significantly
modified. As a specific example, Fig.\,S4 shows the stable and
unstable parameter regimes, by applying the Routh-Hurwitz
criterion \cite{bistability}.

By applying this criterion to the coefficient matrix of these
linear equations, we obtain in the following two stability
conditions
\begin{widetext}
\begin{eqnarray}
S_{1} &=&\left( \omega _{m}^{2}+\frac{\Gamma _{m}^{2}}{4}\right)
\left( \triangle _{2}^{2}+\frac{\gamma ^{2}}{4}\right) -4\omega
_{m}G^{2}\triangle
_{2}>0,   \\
S_{2} &=&\gamma \Gamma _{m}\left[ \left( \triangle _{2}^{2}-\omega
_{m}^{2}\right) ^{2}+\frac{1}{2}\left( \triangle _{2}^{2}+\omega
_{m}^{2}\right) \left( \gamma +\Gamma _{m}\right)
^{2}+\frac{1}{16}\left( \gamma +\Gamma _{m}\right) ^{4}\right]
+4G^{2}\triangle _{2}\omega _{m}\left( \gamma +\Gamma _{m}\right)
^{2}>0,
\end{eqnarray}%
\end{widetext}
where $G=gx_{0}\left\vert a_{2,s}\right\vert $. For $\triangle
_{2}\geq 0$, the second inequality is always satisfied. With the
help of these conditions, the bistability lines can then be
plotted by numerically evaluating the polynomial in Eq.\,(S5). The
resulting figure is shown in Fig. S4, with the corresponding
stable regimes of parameters. Here, to compare with passive COM
systems, we include also optical detunings, ignored previously in
order to focus on the role of optical gain. We see that, in
general, for stronger input power, bistability appears for higher
gain-loss imbalance, as in passive COM systems. In contrast, this
effect can now be completely removed at the gain-loss balance (see
also Ref.\,\cite{transport}).

\end{document}